\makeatletter
\def\input@path{{tex/}}
\makeatother

\documentclass[aps,prl,superscriptaddress,reprint,longbibliography,amsmath,amssymb,onecolumn]{revtex4-2}

\usepackage[ut f8]{inputenc}

\usepackage{natbib}

\usepackage{bm}
\usepackage{graphicx}

\begin{document}
	
	\title{Non-contact acoustic method to measure depth-dependent elastic properties of a kiwifruit}
	
	\author{Laura A. Cobus}
	\email{laura.cobus@auckland.ac.nz}
	\affiliation{Dodd-Walls Centre for Photonic and Quantum Technologies, New Zealand and Department of Physics, University of Auckland, Private Bag 92109, Auckland, New Zealand}
	\author{Kasper van Wijk}
	\affiliation{Dodd-Walls Centre for Photonic and Quantum Technologies, New Zealand and Department of Physics, University of Auckland, Private Bag 92109, Auckland, New Zealand}
	
	\date{\today}
	
	\begin{abstract}
			The mechanical properties of a kiwifruit are investigated via non-contact acoustic measurements. Transmitted ultrasonic waves are measured using a static laser source and a rotating laser ultrasound detector. The measurements enable observation of a  low-frequency surface wave and a higher frequency direct wave; combined with seismology-inspired theoretical modeling, these measurements enable estimation of several elastic parameters for inner and outer fruit flesh layers. Results indicate that the surface and direct wave velocities and the bulk and Young's moduli all evolve with fruit age/ripeness, while only the direct wave velocity and bulk modulus differ significantly between outer and inner fruit layers. 
	\end{abstract}
	
	\maketitle
	
	\section{Introduction}
Non-destructive evaluation of the internal structure of fruit is important for industrial quality assurance and monitoring. Traditionally, the most common parameter used to quantify overall fruit ripeness has been \textit{firmness} -- the force required to deform fruit flesh~\cite{Magness1925}. Firmness is an extensive parameter (dependent on fruit size and shape), but is related to the Young's modulus $E$~\cite{Lu1996,Chen1993} -- a spatially-averaged intensive elastic parameter of the fruit flesh. To measure this and other elastic parameters, acoustic-based  methods are a natural choice, as acoustic  wave propagation is directly related to the mechanical properties of the medium~\cite{Javanaud1988}.

One of the most popular acoustic approaches to date is acoustic resonance spectroscopy, in which vibrations in the fruit are induced
by impact with a small hammer or by a vibrating table, and the resulting modal spectrum is measured with a microphone or Laser
Doppler Vibrometer (LDV). It is now well-established that features of this spectrum correlate with the overall fruit firmness~\cite{Zhang2018}; in addition, this type of measurement can be used to estimate a range of elastic parameters~\cite{Falk1958,Davie1996,Duprat1997,TANIWAKI2009137,Hitchman2016}, each giving slightly different information on fruit flesh properties. Acoustic velocity (measured in the temporal domain) has also been shown to correlate with fruit age and/or firmness~\cite{Self1994,Chen1996,Sugiyama1998,Macrelli2013,Ikeda2015,Hitchman2016,Arai2021}. It remains, however, an ongoing research question as to which parameter may be the most relevant in terms of fruit quality monitoring. The answer may be different for different types of fruit. The quality of fruit may also be strongly affected by its inner structure, motivating research into the properties of individual fruit layers. The majority of such work is destructive, requiring the examination of sections of extracted fruit flesh~\cite{Finney1967,Mizrach1989,Ekrami-Rad2011,Li2012,Kunpeng2017,Du2019,Sakurai2021} or carried out by pushing a probe through the fruit~\cite{Jackson1997}. However, a few studies have exploited the sensitivity of acoustic waves to density variations in the propagation medium for non-destructive evaluation of different flesh layers: ultrasonic surface waves can be used to estimate the properties of the outer flesh layer~\cite{Ikeda2015,Arai2021}, while medical ultrasound imaging approaches can reveal the inner structure of fruit~\cite{Chivers1995,Yoshida2018}. These studies show that while vibrational methods are effective for estimating the average properties of a fruit, and to signal the presence of defects~\cite{Diezma-Iglesias2004,Kadowaki2012,Kawai2018}, acoustic measurements in the temporal domain may have more potential for detailed measurements of inner fruit structure.

\begin{figure}[t]
	\centering
	\includegraphics[width=\textwidth]{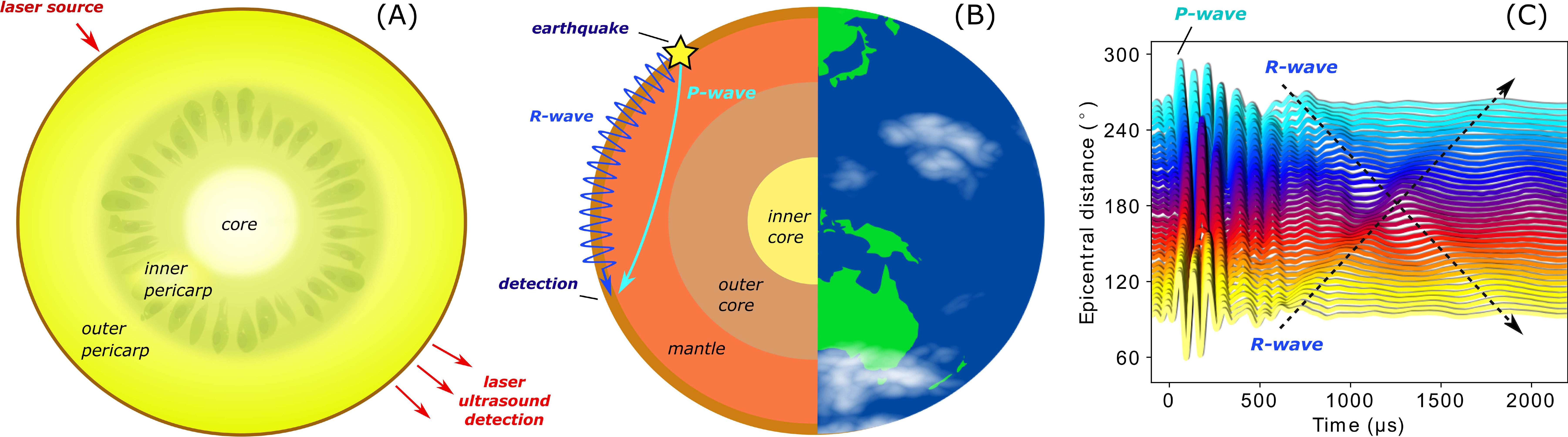}
	\caption{\label{fig:kiwifruit_vs_earth}
		Graphical representation of the analogy between (A) acoustic propagation through the three layers of a kiwifruit, and (B) seismic wave propagation through the three layers of the Earth. (C) Wiggle plot of angle- and time-dependent transmitted acoustic waves through a kiwifruit. At first glance, P- and R-waves appear to be easily identifiable.}
\end{figure}

In this context, we have developed a temporal acoustic approach for the characterization of layered fruit. Here, we demonstrate our method using a kiwifruit. Kiwifruit are the most valuable horticultural export in New Zealand~\cite{NewZealandInstituteforPlantandFoodResearchLimited2020}; however, detailed measurements of their mechanical properties are rare. Parameters which correlate with overall kiwifruit firmness have been measured using acoustic resonance methods with an LDV in detection and with a non-contact electronic waveguide~\cite{Berardinelli2021}; however, kiwifruit are heterogeneous, consisting of a thin outer skin, and three inner fleshy layers: the outer pericarp, inner pericarp, and core~\cite{Ferguson1984} (Fig.\ \ref{fig:kiwifruit_vs_earth}A). Moreover, each of these layers evolves differently as the kiwifruit ages~\cite{Jackson1997,Taglienti2009}. 
Using a penetrometer, Jackson et al. found that, in general, the core is the most
firm, while the inner pericarp is the least firm, most likely because is has a higher liquid content with seed inclusions~\cite{Jackson1997}. More recently, Kenpeng et al. and Du et al. carried out compression testing on extracted portions of kiwifruit flesh to estimate Young's modulus~\cite{Kunpeng2017,Du2019} in the outer pericarp and core.  To our knowledge, these studies -- all of which use destructive testing methods -- have been the only attempts to differentiate between kiwifruit layers. 

Here, we present entirely non-contact and non-destructive measurements of a variety of elastic parameters in kiwifruit flesh.  We record the time- and position- dependent transmitted acoustic waves through a golden kiwifruit (Actinidia chinensis). As the fruit ages, we record multiple sets of these measurements. Previous measurements of elastic parameters on fruit show that some, or all, should correlate directly with firmness, and thus should offer a way to monitor fruit age/quality~\cite{Finney1967,Hitchman2016,Nnodim2019}.
However, such studies are rare, and do not offer clear results in the case of some heterogeneous fruit~\cite{Mizrach2000}. 

In a previous study on apples, we found a remarkable similarity between these datasets and those measured in seismology. We thus used a seismology-inspired methodology to analyze the data, showing that a range of elastic parameters can be estimated~\cite{Hitchman2016}. Here, we apply a similar approach to kiwifruit, with more advanced data acquisition and processing techniques. We find that more detailed and higher quality data offer insights into the  more complex inner structure of kiwifruit. We present an analysis of such a dataset, acquired using a single fruit; as such, this work is not meant to be representative of a large sample of kiwifruit, but rather to be an initial proof-of-concept experiment.  The aims here are to investigate how to interpret these experimental datasets, to determine whether an analysis based on an assumption of homogeneity can be applied to the kiwifruit studied here, and to investigate the potential of this approach for estimating age-related changes in individual kiwifruit layers. 

	\section{Material and Methods}
\label{sec:Methods}

\subsection{Experiment}
\label{sec:methods_experiment}

In our measurements, a high-powered pulsed laser is aimed at one side of the kiwifruit to excite an acoustic waveform in the fruit via non-destructive thermoelastic expansion. On the opposite side of the kiwifruit, a Laser Doppler Vibrometer (LDV) measures any displacement of the surface of the fruit. The LDV is rotated around the equator of
the kiwifruit to measure the surface displacement at multiple locations around the
fruit (Fig.\ \ref{fig:kiwifruit_vs_earth}A). To observe
age/ripeness-related changes in the fruit, these measurements were
repeated multiple times over a time period of 74 hours.

\begin{figure}[ht]
	\centering
	\includegraphics[width=\textwidth]{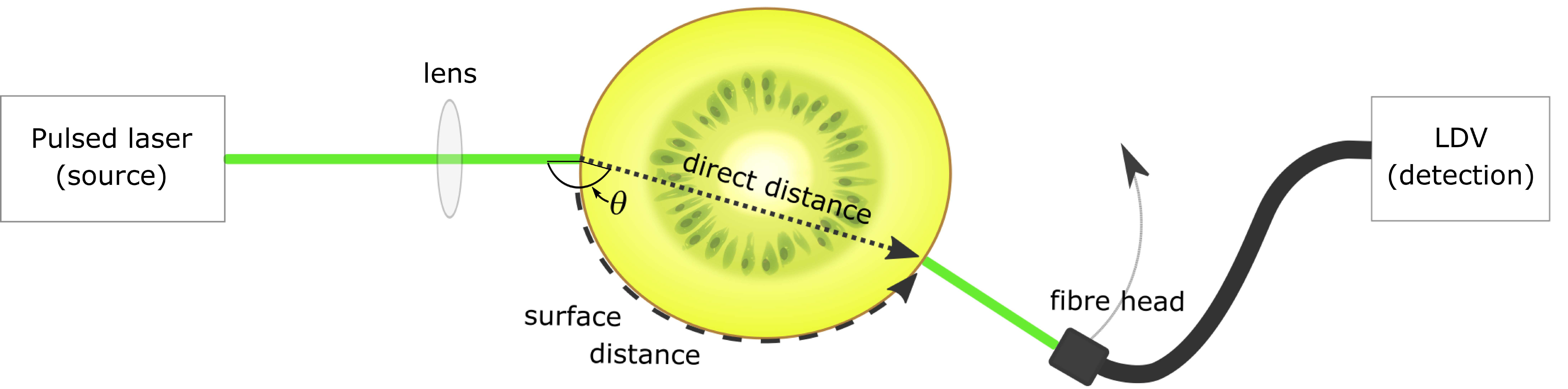}
	\caption{Schematic of experimental setup (not to scale). LDV: Laser Doppler Vibrometer. Also sketched are geometric definitions of the source-detector angle $\theta$, the direct distance $d_\mathrm{direct}$ (dotted line), and surface distance $d_\mathrm{surface}$ (dashed line).}
	\label{fig:setup}
\end{figure}

The experimental setup is sketched in Fig.\ \ref{fig:setup}. The source used to create elastic waves was a Quanta-Ray INDI pulsed laser (central wavelength $1064$~nm, pulse energy $270$~mJ, pulse duration $10$~ns, beam diameter $4$~mm, repetition rate $10$~Hz). The LDV detector was a Polytech OFV-5000Xtra and a MLV-I-120 sensor head, with a flexible fibre optic cable and lens attached. The detection beam could be focused through the cable and lens to a point of $3$~mm
on the fruit surface. Reflective tape was applied around the fruit equator to optimize the amount of detected light (note that golden kiwifruit have a smooth, hairless surface, enabling good adhesion with the reflective tape). Signals were recorded by an Alazar Tech digital oscilloscope card at $1$~MS/s and 16-bits dynamic range. The entire data acquisition system was controlled by the open-source software PLACE (Python package for Laboratory Automation, Control, and Experimentation)~\cite{Johnson2015}. We have previously reported the approach described above to study apples~\cite{Hitchman2016}, with the exception of the detection sensor
head which was placed on a long rotating arm. Now, with the sensor head attached to a fibre optic cable, the data collection is much more flexible, reliable and rapid.

The golden kiwifruit studied was purchased from a local grocery store. As with most golden kiwifruit available in New Zealand, the cross-section of the fruit is elliptical.  In our experiment, both major and minor radii were measured using calipers, and this geometry taken into account in subsequent data analysis. The fruit was positioned such that the maximum source-detector distance ($\theta=180^\circ$) is the maximum direct distance through the fruit, i.e. twice the major radius. Data was recorded at seven different times spanning a time lapse of 74 hours, beginning with the day the fruit was purchased. For each data acquisition, the fibre head was rotated around the fruit from $\theta\approx 90^\circ-280^\circ$, stopping at intervals of $10^\circ$ to record the transmitted time-dependent longitudinal waves. Between acquisitions the fruit was stationary. Fruit volume was measured using Archimedes' principle; both volume and weight were measured before and after the entire time lapse, and extrapolated values were calculated for the other acquisition times, assuming a linear rate of change between first and last points. From these points, density $\rho$ was estimated.

	\subsection{Analysis and interpretation}

Using the approach described previously, 
each dataset acquired is a matrix of transmitted acoustic amplitude $R(\theta,t)$, where $t$ is the amount of time recorded by the detector after the source excitation, and \textit{epicentral distance} $\theta$ is the angle between source and detector points. This type of dataset, often referred to as the \textit{common-source gather} in exploration seismology, is directly analogous to that acquired in seismology, in which multiple detectors on the surface of the Earth detect vibrations resulting from a single seismic event (Fig.\ \ref{fig:kiwifruit_vs_earth}B).  Figure\
\ref{fig:kiwifruit_vs_earth}C shows a visualization of $R(\theta,t)$.  Referred to as a \textit{wiggle plot} in seismology, this type of figure facilitates distinction between the different types of waves which contribute to the total transmitted wavefield~\cite{Chivers1995,Stein2009}. In Fig.\ \ref{fig:kiwifruit_vs_earth}C, two different waves can be observed; the most straightforward interpretation of these waves is that they are (\textit{i}) a direct compressional wave, called the primary or `P-wave' in seismology, and (\textit{ii}) surface Rayleigh waves travelling around each side of the fruit, called the `R-wave'~\cite{Chivers1995,Hitchman2016}. For each wave, of interest is the arrival time of the respective wave at each detector position. These {\it arrival times} were determined using the following stpdf: (\textit{i}) first-arriving signals were manually picked for traces recorded at $\theta=180^\circ$, (\textit{ii}) temporal cross-correlation of signals recorded at different source-detector angles was used to determine the relative arrival times for the range of detector positions, and (\textit{iii}) these times were corrected by the first arrival time at $\theta=180^\circ$. This process was performed for both the P- and R-waves. 
	
	\subsubsection{Estimating acoustic wave velocity}
\label{sec:methods_analysis_v}

The velocity of Rayleigh waves travelling along the surface of the kiwifruit is given by $v_R = \Delta d_\textrm{surface}/\Delta t_R$, where $t_R$ is arrival time and $d_\textrm{surface}$ is the distance traveled around the fruit surface.  The surface distance $d_\textrm{surface}$ is sketched in Fig.\ \ref{fig:setup}. For a kiwifruit, $d_\textrm{surface}$  can be defined as the arc length of an ellipse, and can be calculated from the minor and major radii. Thus, $v_R$ can be measured by performing a linear fit of  $d_\textrm{surface}$ versus the experimentally-measured $t_R$~\cite{Hitchman2016,Cobus2020}.
 
Care must be taken in estimating $v_R$, as near $\theta=180^\circ$ the data contains a superposition of the two R-waves that have travelled around the epicenter of the fruit in opposite directions -- an effect that can be seen in Fig.\ \ref{fig:kiwifruit_vs_earth}C. This impedes an accurate measurement of R-wave arrival time near $\theta=180^\circ$. To estimate $v_R$, therefore, these points are excluded from the linear fit.

If the fruit is homogeneous, i.e. if acoustic wave velocity is constant throughout the fruit, then the same approach can be taken for P-waves; arrival time $t_P$ should scale linearly with direct distance $d_\textrm{direct}$ through the fruit, where $d_\textrm{direct}$ is defined as the straight-line path between source and detector (Fig.\ \ref{fig:setup}).   
Then, P-wave velocity can be found by fitting the data with a linear fit: $v_P=\Delta d_\textrm{direct}/\Delta t_P$.  However, the heterogeneous nature of kiwifruit (e.g. Fig.\ \ref{fig:kiwifruit_vs_earth}B) means that this linear relationship is not correct. For a more accurate model of how the P-wave propagates through a  heterogeneous fruit, we used the TauP Toolkit~\cite{Crotwell1999}. This java software was originally designed for seismological ray-racing; here, it is employed to simulate compressional wave propagation through a layered fruit.  To begin, a guess for the depth-dependent velocity profile is defined. The toolkit then simulates wave propagation through a sphere with such a velocity profile, and predicts the resulting wave arrival times at points along the surface (details on this simple process are given in Ref.\ \cite{TauPmanual}). By comparing these arrival time predictions with experimental measurements, the velocity model can be optimized, meaning that in principle, an estimate can be obtained of $v_P$ for each layer of the kiwifruit. For this study, we performed this optimization manually.

	\subsubsection{Estimating other elastic parameters}

If both $v_R$ and $v_P$ are known, a range of elastic parameters
can be estimated for the kiwifruit.  While the experimental setup used
here can not detect shear waves (S-waves) directly (the LDV only
measures the longitudinal component of the kiwifruit surface
displacement), shear-wave velocity $v_S$ can be estimated from $v_R$
and $v_P$ by finding the real roots of
\begin{equation}
\label{eq:vS}
\left( 2-\frac{v^2_R}{v^2_S}\right)^2+4\sqrt{\frac{v^2_R}{v^2_S}-1}\sqrt{\frac{v^2_R}{v^2_P}-1}=0,
\end{equation}
and requiring that $0<v_R<vS$.
Elastic moduli can then be calculated~\cite{Stein2009}:
\begin{align}
	\label{eq:E}
	E &= v_S^2\rho\left(2\nu+2 \right),\\
	\label{eq:G}
	G &= E/\left[2(1+\nu)\right], \textrm{ and}\\
	\label{eq:K}
	K &= E/\left[3(1-2\nu)\right], 
\end{align}
where $E$ is Young's modulus, $G$ is the shear modulus, $K$ is the bulk modulus, and
\begin{equation}
\label{eq:nu}
\nu\equiv \frac{1}{2}\frac{\left(v_P^2/v_S^2\right)-2}{\left(v_P^2/v_S^2\right)-1}
\end{equation}
is Poisson's ratio.
The elastic moduli measure the response of the material to applied
deformation: Poisson's ratio quantifies the deformation of a material
perpendicular to the direction of an applied force, Young's modulus is
the ratio of stress to strain, and thus measures the stiffness of the
material, the shear modulus measures the material stiffness with
respect to \textit{shear} deformation, and the bulk modulus is related
to the inverse of compressibility, representing the change in volume
to external stress.

\section{Results and Discussion}

\begin{figure}[ht]
	\centering
	\includegraphics[width=0.8\columnwidth]{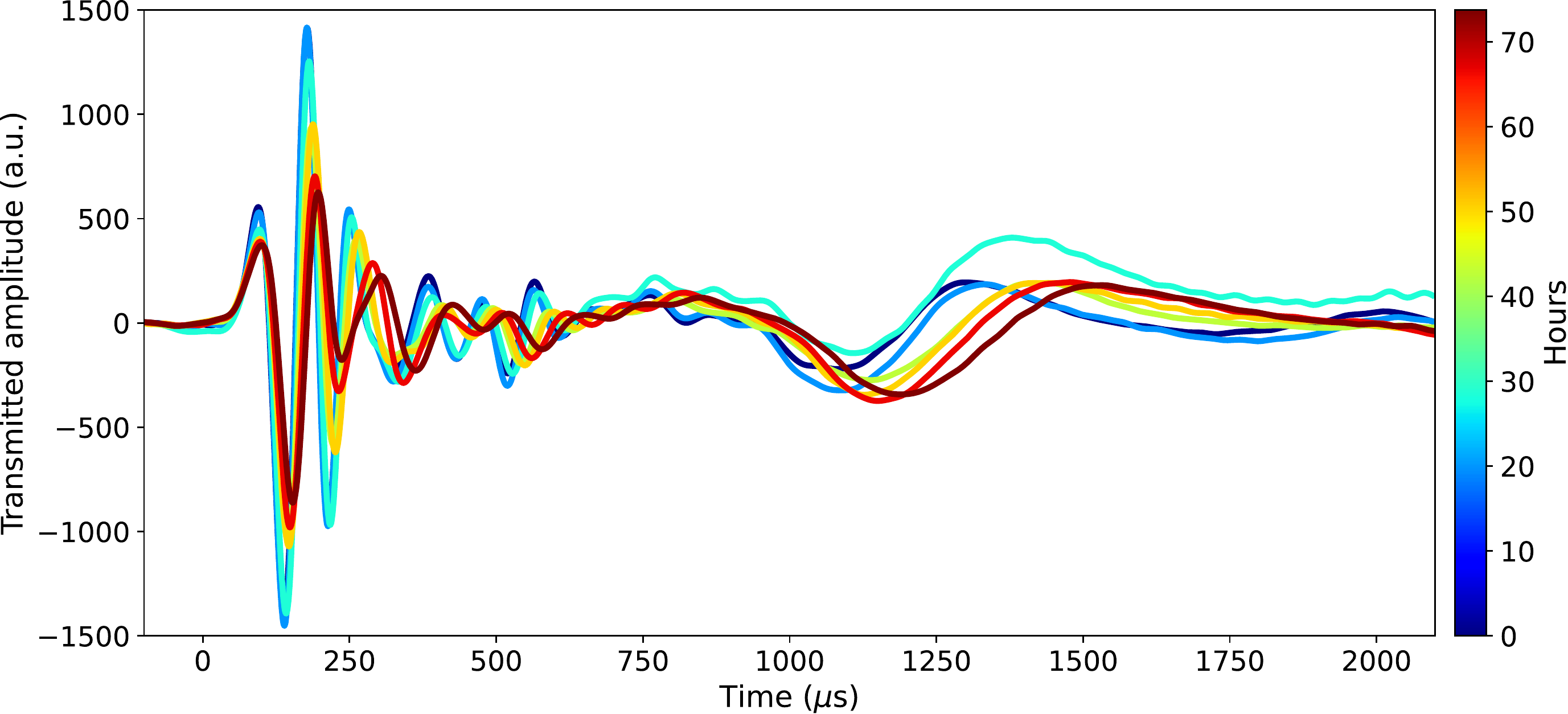}
	\caption{Transmitted acoustic waves through a golden kiwifruit at $\theta=180^\circ$, measured periodically over a range of 74 hours. } 
	\label{fig:Trans180}
\end{figure}

Figure\ \ref{fig:Trans180} shows transmitted wavefields through a kiwifruit at epicentral distance $\theta=180^\circ$, measured at seven different time lapses over a 74-hour period. 
The first-arriving P-wave ($t\sim50-250$~$\mu$s) and the later-arriving R-wave ($t\sim1000-1300$~$\mu$s) arrive later in time as the fruit ages. Frequency filtering can be used to isolate each wave from the recorded dataset; the early-arriving waves are in the range $f\approx 5-16$~kHz while later arrivals are in the
lower range $f\approx 1-4$~kHz.

\begin{figure}[b]
	\centering
	\includegraphics[width=0.9\columnwidth]{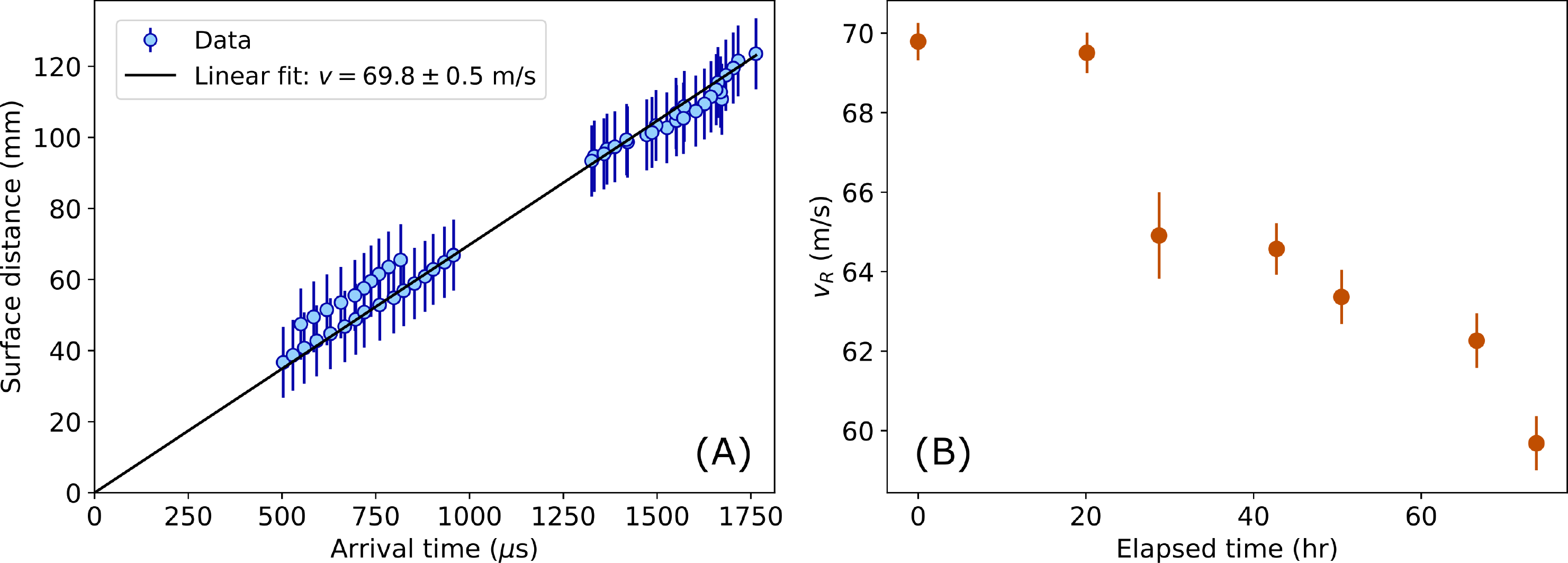}
	\caption{(A) Experimentally-measured R-wave arrival times are
		plotted versus $d_\textrm{surface}$, distance travelled along the
		fruit surface from the source point (symbols). The slope of a
		linear fit (black line) gives an estimate of velocity
		$v_R$. Experimental data shown was measured after a time-lapse of
		0 hours. (B) Results for $v_R$ are shown for the entire time lapse
		period.}
	\label{fig:vR_linearfit}
\end{figure}

Figure\ \ref{fig:vR_linearfit}A shows a representative example of how the R-wave velocity, $v_R$, is estimated via a linear fit of distance travelled along the surface, $d_\textrm{surface}$, versus arrival time $t$. As discussed, 
points near $\theta=180^\circ$ contain a superposition of two R-waves, and are not included in the linear fit. In Fig.\ \ref{fig:vR_linearfit}, arrival times from waves circling the fruit in both directions (see Fig.\
\ref{fig:kiwifruit_vs_earth}C) are shown, causing the points to appear to overlap. Results for the velocity extracted from the linear fitting are shown in Figure\ \ref{fig:vR_linearfit}B over the entire time lapse period, showing that as the fruit ages/ripens, $v_R$ decreases.

\subsection{Interpretation of the first-arriving wave}

\begin{figure}[ht]
	\centering
	\includegraphics[width=0.9\columnwidth]{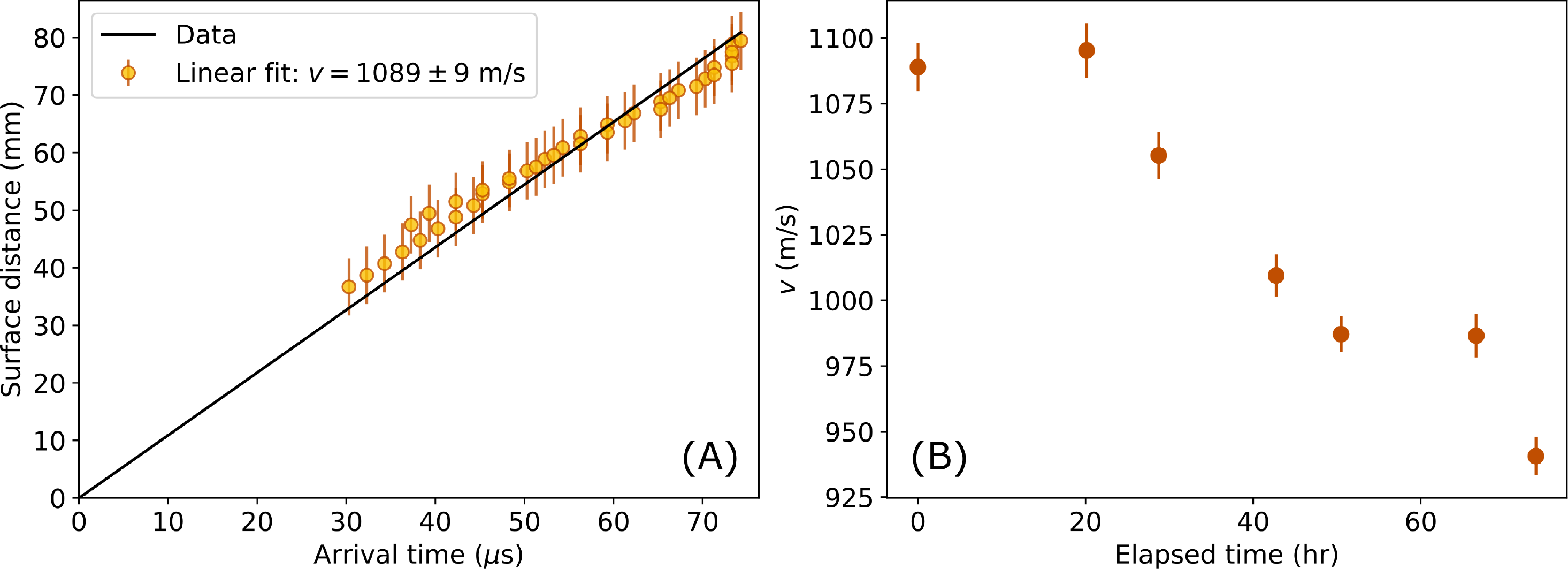}
	\caption{(A) Experimentally-measured arrival times of the
		first-arriving wave are plotted versus $d_\textrm{surface}$,
		distance travelled along the fruit surface from the source point
		(symbols). Supposing that this wave is a surface wave, the slope
		of a linear fit (black line) gives an estimate of velocity
		$v$. Experimental data shown was measured after a time-lapse of 0
		hours. (B) Results for $v$ are shown for the entire time lapse
		period.}
	\label{fig:vP_linearfit}
\end{figure}

As discussed previously, the first-arriving wave is typically identified as a P-wave, or `direct' wave -- a compressional wave that has traveled through the bulk of the material. For a homogeneous fruit, the P-wave arrival times $t$ would scale linearly with the direct path through the fruit from source to detector, $d_\text{direct}$, giving the P-wave velocity $v_P=D_\text{direct}/t$. However, this is not what is observed; as shown in Fig.\ \ref{fig:vP_linearfit}A, the arrival times correlate more strongly with the distance along the surface, $d_\text{surface}$. 
This observation is similar to that in a study by Arai et al., in which two waves propagate in mango flesh at different speeds, both with a roughly linear relationship between angle and arrival time~\cite{Arai2021}. The interpretation given by Arai et al. is that both are surface waves, where one is confined to the skin, the other to the outer flesh layer, and with $v_\mathrm{skin}\sim 10 v_\mathrm{flesh}$. In this interpretation, a simple linear fit of $d_\textrm{surface}$ vs $t$ should give the velocity of the first-arriving surface wave. Figure\ \ref{fig:vP_linearfit}A shows that our data can also be described by this simple model. Results for $v_P$ extracted from the linear fit are shown in Figure\ \ref{fig:vP_linearfit}B over the entire time lapse period; as observed with $v_R$, the estimated $v_P$ decreases with fruit age.

The interpretation of Arai et al. is convincing, in their case, due to the thickness of the outer mango flesh layer, and to the presence of the large, hard pit in the mango that prevents the arrival of a direct compressional wave unlikely. 
However, previous measurements on apples using using a similar measurement technique as that described here have shown clear evidence that the first-arriving wave is a P-wave~\cite{Hitchman2016}. While apples and kiwifruit do have significant differences in structure and mechanical properties, it is reasonable to suppose that a direct wave should be observable for kiwifruit as well, given the softness of kiwifruit and the high sensitivity of the LDV detection method employed. 
In a numerical study on acoustic wave propagation in orange peels, Jimenez et al. also observe a first-arriving wave with a linear relationship between angle and arrival time, and also interpret it as a compressional P-wave~\cite{Jimenez2012}. Thus, the more likely explanation for our observations is that the first-arriving wave is a compressional P-wave whose propagation is influenced by the heterogeneity of the kiwifruit. This argument is supported by slight biases visible in the linear fit of Fig.\ \ref{fig:vP_linearfit}A: for short distances, the estimated travel time is consistently later than the linear fit, while for large distance the estimated travel time is consistently earlier than the fit.

\subsection{Comparison of P-wave arrival times with seismic ray-tracing model}
\label{sec:comparison_expsim}

\begin{figure}[h]
	\centering
	\includegraphics[width=0.95\columnwidth]{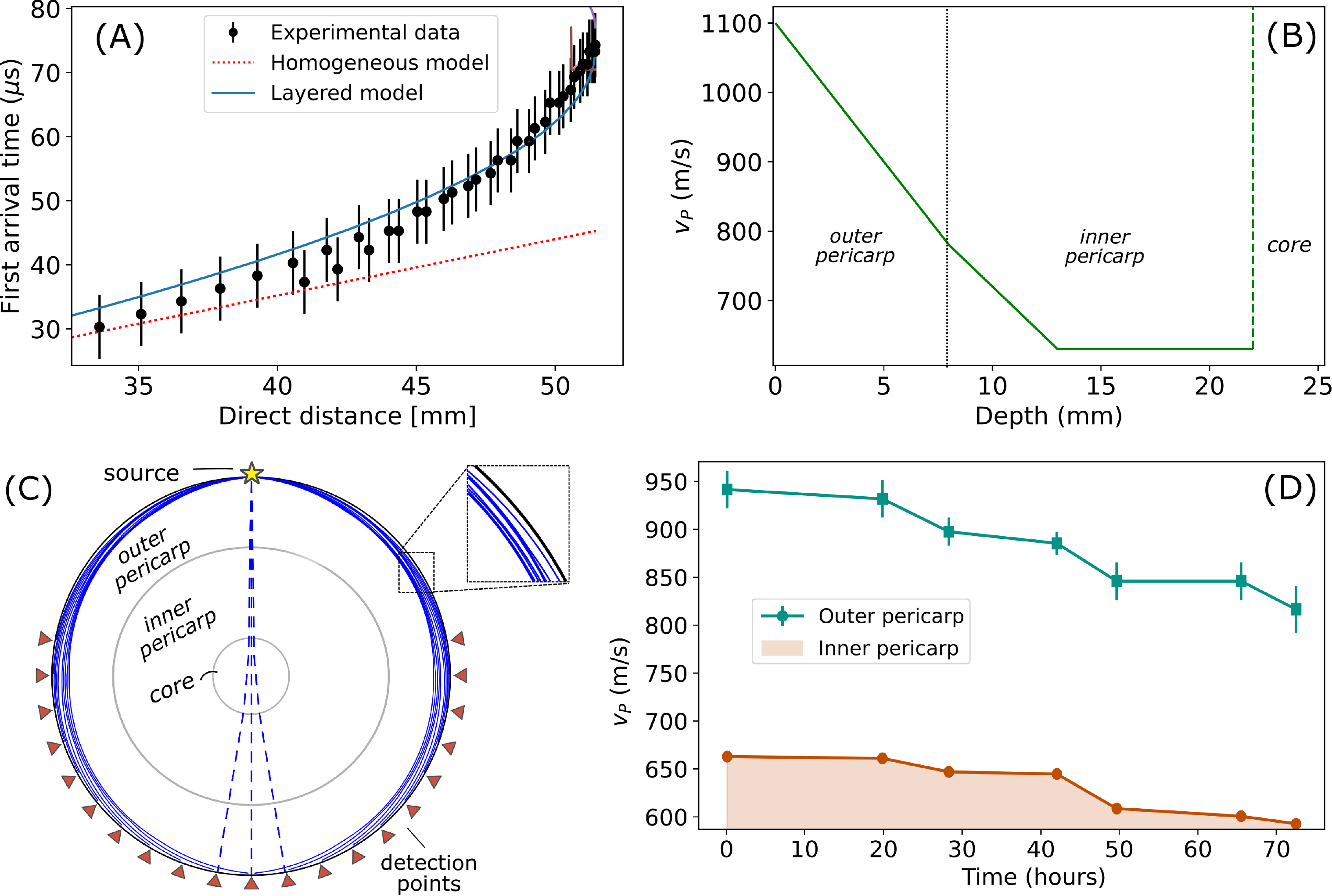}
	\caption{(A,B) Experimental P-wave arrival times $t_P$ (symbols), acquired at a time lapse of $0$~hours, are compared with theoretical predictions from homogeneous (red dotted line) and layered fruit models (green solid lines). $t_P$ is plotted versus (A)  $d_\textrm{direct}$ (to emphasise the deviation from the homogeneous model. (B) The layered velocity model used to predict $t_P$ in (A,B). Values in the core (depth~$>23$~mm) do not impact the predictions, and are indicated with a dashed line. (C) Ray paths (blue lines) corresponding to the first-arriving P-wave for the layered model. A zoom of the top right corner shows the individual paths of each ray. Dashed lines indicate secondary arrivals which lie within the experimental uncertainty. For clarity, only every second detector position is shown. (D) Estimates of $v_P$ in the outer pericarp (teal circular symbols) and inner pericarp (orange square symbols), for each time-lapse measurement. The shaded orange area indicates that the data points constitute an upper bound for $v_P$ in the inner pericarp.}
	\label{fig:vP_arrival_times}
\end{figure}

Figure\ \ref{fig:vP_arrival_times}A shows the arrival times of the P-wave versus direct distance $d_\text{direct}$. The data can not be described by a homogeneous model, as indicated by the red dotted line; thus, a a more realistic realistic heterogeneous model for the fruit is required.  Figure \ref{fig:vP_arrival_times}B shows a simplistic but realistic velocity model for a heterogeneous kiwifruit. The three main layers are represented (Fig.\ \ref{fig:kiwifruit_vs_earth}A): the outer pericarp (depth $\sim 0-12$~mm), with a velocity gradient to the inner pericarp (depth $\sim 12-23$~mm), and the smaller core (depth $\sim 23-26$~mm). The general form of this model resembles results from experimental measurements by Jackson et al. of depth-dependent firmness in other types  of kiwifruit~\cite{Jackson1997}. Here, however, we do not include the skin in our model, as the range of wavelengths of our measurements (on the order of centimetres) are likely too large to be sensitive to such a thin layer.

To compare the velocity model hypothesis with experimental results, we simulate the propagation of P-waves through the velocity model using a ray-tracing method which was initially developed to model elastic wave propagation in the Earth. 
Representative results are shown in Fig.\ \ref{fig:vP_arrival_times}.  For most source-detector angles, the theoretically-predicted arrival times agree with experimental measurements (Fig.\ \ref{fig:vP_arrival_times}A). It is of note that the data is not symmetric about $\theta=180^\circ$; this effect is likely due to a slight asymmetry in the fruit shape, and also observed in the R-wave arrival times.  The theoretical simulation also gives the paths that the compressional waves take through the velocity model.  Figure\ \ref{fig:vP_arrival_times}C shows the ray paths associated with the simulated arrival times of Fig.\ \ref{fig:vP_arrival_times}A. For this velocity model, the P-waves which arrive first to most detectors are those which have travelled around the outer pericarp and the gradient layer between the outer and inner pericarp. 
This `bending' of the compressional P-waves is due to the gradient in $v_P$ (Fig.\ \ref{fig:vP_arrival_times}B), which causes the P-wave to refract according to Snell's Law away from a straight line (towards the fruit centre). It is of note that the simulation also predicts a few secondary arrivals which have travelled through the centre of the fruit (dashed blue lines in Fig.\  \ref{fig:vP_arrival_times}C). Although rays following these paths arrive after those which were confined to the outer pericarp, they still lie within the experimental uncertainty (the error bars of Fig.\  \ref{fig:vP_arrival_times}A), and thus we can not make a definitive statement about which path the waves have taken to arrive at epicentral distances $\theta=160^\circ-200^\circ$. This interesting question will be investigated in more detail in future studies. \\

Each of the 7 datasets recorded within the 74-hour time-lapse period show the same approximate scaling of arrival time with distance seen in Fig.\ \ref{fig:vP_arrival_times}A. Thus, the ray-tracing simulation and comparison to data was performed for each dataset. 
In this way, estimates for $v_P$ were obtained for the outer two layers.  It is important to note that a very wide variety of velocity models were tested, and none predicted arrival times which agree with the experimental data except those reported here (e.g. Fig.\ \ref{fig:vP_arrival_times}B).  Results for $v_P$ are shown in Fig.\ \ref{fig:vP_arrival_times}D. 
While we can not make a definitive measurement of the core velocity (as discussed in the previous paragraph), we find that a specific negative velocity gradient between the outer and inner pericarp is required for the simulation and experimental data to agree.  Thus, we can estimate $v_P$ in the outer pericarp, and set an upper bound on $v_P$ for the inner pericarp.
The best-fitting model (Fig.\ \ref{fig:vP_arrival_times}B) predicts that the average P-wave speed in the outer pericarp (for a time lapse of $0$~hours) is $v_P\sim 945$~m/s. This result is similar to $v= 1089\pm9$~m/s -- the velocity obtained by interpreting the first-arriving wave as a high-frequency surface wave (Fig.\ \ref{fig:vP_linearfit}B). However, while the surface-wave interpretation gives an estimate of acoustic velocity near the fruit surface, the P-wave interpretation (and subsequent ray-tracing modeling) allows limits to be set for $v_P$ in both the inner and outer pericarp, and for the velocity gradient between them (Fig.\ \ref{fig:vP_arrival_times}D).

\subsection{Elastic parameters of the inner and outer pericarp}

\begin{figure}[ht]
	\centering
	\includegraphics[width=\columnwidth]{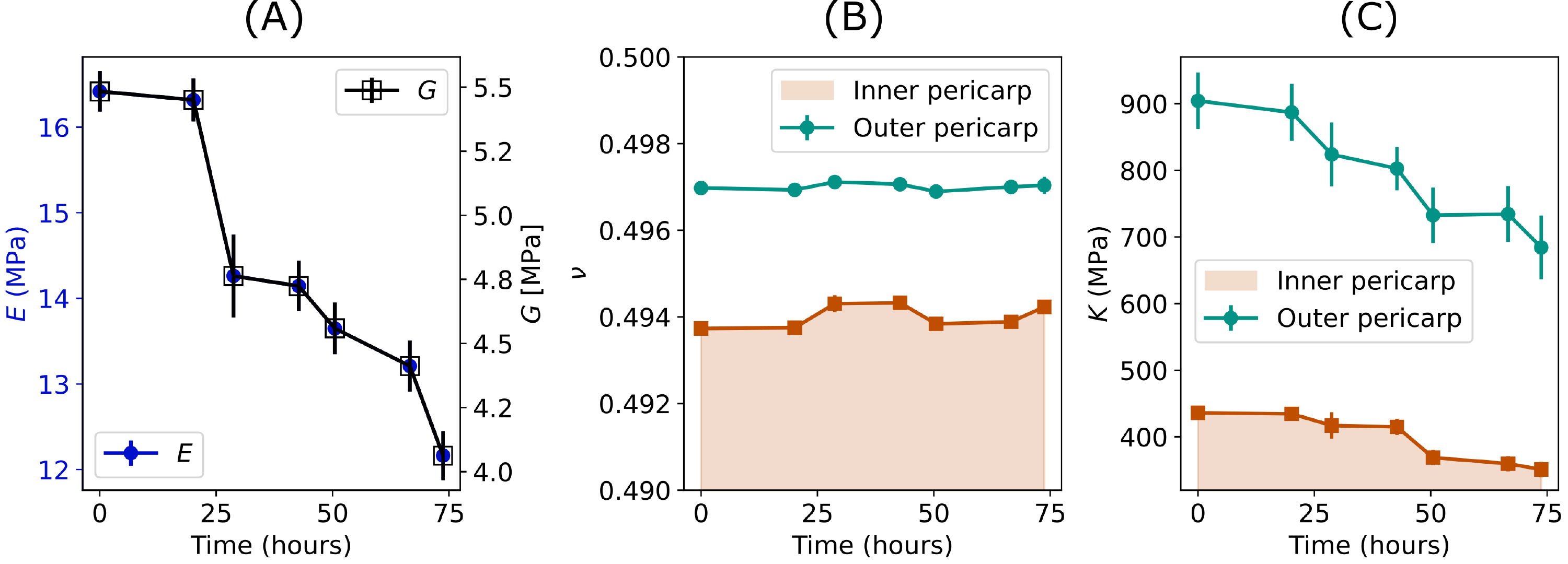}
	\caption{Elastic wave parameters of a golden kiwifruit, measured
		over a time period of 74 hours. In (A) $E$ and $G$ in the outer
		pericarp are shown (values in the inner pericarp are identical,
		but constitute upper bounds). Values for (B) $\nu$ and (C) $K$ are
		shown for the outer pericarp (teal circular symbols) and inner
		pericarp (orange square symbols). }
	\label{fig:elasticparams_timelapse}
\end{figure}

Using Eqns.\ \ref{eq:vS}-\ref{eq:nu}), Poisson's ratio $\nu$ and elastic moduli $E$, $G$, and $K$ were calculated for the outer and inner pericarp separately. Over the time lapse period, density 
increased very slightly from $\rho=1.027\pm 0.005$~g/cm$^3$ to $\rho=1.040\pm 0.005$~g/cm$^3$. The relative long wavelength of the Rayleigh wave velocity $v_R$ compared with the kiwifruit size means that the R-wave likely samples both inner and outer pericarps and the core, with $v_R$ being the average velocity of these layers. 
Thus, the same $v_R$ value is used in Eq.\ \ref{eq:vS} for both pericarp layers. 
Results for selected elastic parameters are shown in Fig.\
\ref{fig:elasticparams_timelapse}.  We find that while estimates of $E$ and $G$ do not vary significantly between the inner and outer pericarp, they do exhibit a clear decay with fruit age (Fig.\ \ref{fig:elasticparams_timelapse}A,B). (Note that values for the inner pericarp are almost identical, but constitute an upper bound.)

Conversely, while $\nu$ varies with layer, it does not evolve with fruit age (Fig.\ \ref{fig:elasticparams_timelapse}B). The bulk modulus $K$ is the only elastic parameter to change with both time and spatial location in the fruit (Fig.\ \ref{fig:elasticparams_timelapse}C).

\subsection{Comparison with previous results for kiwifruit and other fruit}

Most previous studies of acoustic velocity in fruit do not discriminate between bulk and surface waves; thus, comparison of our results for acoustic velocity with previous measurements must done be with care.
In 1997, Muramatsu et al. measured acoustic travel time through a green kiwifruit using two transducers in contact with either side of the fruit. While they interpreted the measured wave as a bulk wave, their excitation was on the order of Hz, and the observed wave arrives after $1$~ms, leading us to believe that they only observed the Rayleigh wave discussed in this paper, not the P-wave. Indeed, we have reproduced this experiment using two transducers with a higher-frequency excitation (not shown), and we observed both the P- and R-waves described in this paper. More recently, shockwave-induced Rayleigh waves were used to measure $v_R\sim 25-50$~m/s in the outer flesh of mangos,  finding that $v_R$ decreases as the mango ripens~\cite{Arai2021}. This measurement agrees with previous measurements of a ''transmission velocity'' by Sugiyama et al.~\cite{Sugiyama1998}. Our values of $v_R$ are higher, but on the same order, and the same trend with fruit age is observed, indicating that $v_R$ could be a useful basic observable for predicting overall kiwifruit age/ripeness. 

Measurements of $v_P$ for fruit are much more rare than $v_R$. Our measured values for $v_P$ are higher than those reported for apples ($\sim 182$~m/s)~\cite{Hitchman2016}, 
and are, in fact, closer those measured for potatoes ($v\sim 500-800$~m/s)~\cite{Cheng1994}.

\subsubsection{Elastic parameters}

Young's modulus $E$ may be the closest parameter to that which is evaluated by the customer who squeezes a fruit to assess its quality. To our knowledge, $E$ is the only elastic parameter previously measured for kiwifruit; for green kiwifruit (Actinidia deliciosa), $E\sim 1-5$~MPa~\cite{Kunpeng2017,Pourkhak2017} and for red-fleshed gold kiwifruit (Actinidia chinensis cv. Hongyang), $E$ varied from $E=2.389\pm 0.545$~MPa to $E=0.288\pm 0.064$~MPa, measured six days later~\cite{Du2019}. This decrease of $E$ with fruit age agrees with our observations (Fig.\ \ref{fig:elasticparams_timelapse}B). Our values of $E\sim 9-12$~MPa are of the same order of magnitude as the previous values, but consistently larger. 

For other types of fruit, a very wide range of values of elastic
parameters have been reported. Our values for the shear modulus in the kiwifruit outer pericarp ($G\sim 3-4$~MPa) are on the order of estimates for other fruit:  $G<1$~MPa for pomegranates~\cite{Ekrami-Rad2011} and for banana and apple flesh~\cite{Sakurai2021} and $G\sim 5-6$~MPa for pear flesh~\cite{Finney1967}.
For Poisson's ratio, $\nu\sim 0.16-0.24$ for
apples~\cite{Hitchman2016}, $\nu\sim 0.03-0.4$ for apple
flesh~\cite{Finney1967,Sakurai2021}, $\nu\sim 0.408$ for banana
flesh~\cite{Sakurai2021}, and $\nu\sim 0.25-0.4$ for pear
flesh~\cite{Finney1967}. We find relatively high values, $\nu>0.49$
for kiwifruit. 

Very few reports of experimental measurements of the Bulk modulus $K$
for fruit are available: $K\sim 0.4-0.8$~MPa has been reported for
banana flesh~\cite{Sakurai2021}, $K\sim 1.5-3.1$~MPa for apple
flesh~\cite{Sakurai2021}, and $K\sim 3-7$~MPa for
peaches~\cite{Clark1977}. Our values of $200-900$~MPa for golden
kiwifruit are much larger.  \\

There are several possible reasons for the fact that our values of $E$, $K$ and $\nu$ are larger than might be expected. For $E$, kiwifruit previously studied were of a different type. The kiwifruit studied here was also very firm when measurements began (as evaluated by feel), and was thus perhaps less compressible than other fruit
studied. It also possible that cold storage of the fruit before it reached the grocery store increases its compressibility~\cite{Nicolas1986}.  In general, however, we can
expect elastic constants measured via acoustic techniques to be higher than those using quasi-static stress/strain tests. This effect has been observed for apples~\cite{Finney1967,Varela2007,Hitchman2016}, orange peel~\cite{Jimenez2012} and watermelons~\cite{Ikeda2015}. Acoustic methods measure the `dynamic' Young's modulus $E_d$, which is similar to the true $E$, while stress/strain-type approaches measure the `apparent' Young's modulus $E_a$, which can be lower than $E_d$, especially if the material being tested becomes close to deformation. These differences are due to the
frequency dependence~\cite{Varela2007,Hitchman2016} of the elastic fruit parameters.  The loss of information inherent in low-frequency static measurements means that parameters which are sensitive to $v_P$, a higher-frequency wave, could be underestimated~\cite{Jimenez2012}. In addition, acoustic measurements examine the properties of a fruit whose microstructure is unchanged during the experiment, as opposed to deformation experiments which may be closer to examining nonlinear behaviour of the flesh. If the fruit microstructure is even slightly crushed, the value of $\nu$ should decrease, with a related change in the other elastic parameters as well.  It is possible that fruit with higher water content are more likely to exhibit this effect, having higher values for (high-frequency) bulk compressional waves, most likely a larger difference between $v_R$ and $v_P$, and a larger resulting value of $\nu$. In that case, parameters such as $K$ which are more sensitive to changes in $\nu$ (as opposed to the relatively weak dependence of $E$ on $\nu$) would differ even more drastically from those estimated by low-frequency or quasi-static measurements.\\

\section{Conclusion and perspectives}

We have reported the first dynamic measurements of kiwifruit mechanical properties. Using an entirely non-contact experimental approach, we measured different types of ultrasonic waves propagating in the fruit. We also observed a deviation from the expected behaviour of the first-arriving wave; using theoretical modeling inspired by seismological techniques, we find that the first-arriving wave can be interpreted as a compressional wave which is confined to the outer pericarp (the outermost part of the kiwifruit flesh). 

This analysis enables us to estimate several elastic parameters for the outer pericarp, and limits on these parameters for the inner pericarp. Considering previous static or frequency-limited measurements for a kiwifruit, we observe comparable but slightly higher values of $E$ than previously reported. We also present the first estimates of $K$ for a kiwifruit, which we find to be much higher than those estimated for different types of fruit. These differences may be attributed to the different physical properties measured using dynamic as opposed to static measurements. Finally, we find that many of the elastic parameters estimated are sensitive to fruit age. The simplest observable with which to track kiwifruit ripening is $v_R$, the velocity of the low-frequency surface wave. However, a more detailed examination of the P-wave propagation seems promising for probing the fruit's inner structure, and the evolution of each kiwifruit layer with ripening.

With this proof-of-concept experiment, we find that dynamic elastic moduli for individual kiwifruit layers can be estimated non-invasively. Previous work has established that these layers evolve differently with age~\cite{Jackson1997,Taglienti2009} and are affected differently by disorders such as cold storage breakdown (a disorder in which cell walls break down over time)~\cite{Burdon2011}. Our technique could thus be valuable in the future for industrial monitoring of kiwifruit quality. Our all-optical experimental approach could be scaled down and optimized for a rapid acquisition and automated analysis of the datasets presented in this article. However, more research is required to enable such a technology. A current limitation is that relatively high source laser power was used in order to optimize signal to noise for easier distinction between P- and R-waves. Over time, this could potentially result in some damage to the fruit at the source point. Future experiments will investigate the lower limits for source power; alternate methods of non-contact excitation~\cite{Hosoya2017,Arai2021} may be required to remove the possibility of ablation altogether. Another limitation to rapid measurements on multiple fruit is that the fruit orientation relative to the source/receiver must be known with accuracy. This limitation could be bypassed if the fruit orientation could be determined from the arrival times of surface waves, or by using another integrated technology to measure the fruit geometry. More accurate knowledge of the outer fruit shape would also reduce the uncertainty in velocity measurements, which currently assume a perfectly elliptical cross-section. Another drawback in this work was the manual optimization of the comparison between experimental and simulated results. Additional accuracy and automation of the modeling of acoustic propagation through a layered kiwifruit could be achieved via comprehensive fitting of the data with a set of theoretical predictions, calculated over the entire space of possible velocities and layer boundary positions. We are currently working to develop such a fitting procedure. Finally, it will be necessary to conduct experimental trials on multiple fruit for a better estimate of typical elastic parameters for golden kiwifruit.

Other future perspectives include analysis of the attenuation of each type of wave, and of the frequency-dependence of acoustic velocities and elastic parameters as the fruit ages. Both of these approaches could give added information and precision in tracking fruit quality. More generally, our use of ray-tracing for characterizing spherical layered media may be useful for other non-destructive applications~\cite{Amziane2012,Wang2020}.

\bibliography{main_bib}
	
\end{document}